\documentstyle[editedvolume]{crckapb} 

\begin{opening}
\title{OPTICAL ASTRONOMY IN POST-APARTHEID SOUTH AFRICA:\protect\\
       1994 TO 2004}

\author{PATRICIA WHITELOCK}
\institute{South African Astronomical Observatory\\
           PO Box 9, Observatory, 7935, South Africa\\
	   paw@saao.ac.za}

\end{opening}

\runningtitle{ASTRONOMY IN SOUTH AFRICA}

\begin{document}

\begin{abstract}
 The progress of optical astronomy in post-apartheid South Africa is
discussed. Particular emphasis is given to the socio-political climate which
embraced the idea of a 10-m class telescope as a flagship project that would
lead to widespread development in science, technology and education - not
only in South Africa, but across the subcontinent. 
\end{abstract}

\section{Introduction}
 This account of optical astronomy in South Africa starts where Feast (2002,
hereafter Paper I) left off, in 1994 with the first democratic elections and
the start of a new era. The end of apartheid offered vastly increased
opportunities for international collaborations among individuals and
institutions, which the astronomy community was quick to take advantage of.
Nevertheless, while the historical strength of astronomy laid a firm
foundation for growth and success, the reasons why the discipline 
thrived and grew in the following decade were complex and essentially
political.

 In the following I attempt a brief description of the policies and
socio-political climate that have nurtured astronomy in South Africa, while
acknowledging that no two individuals will see this in the same way and that
any such account will be incomplete and probably idiosyncratic.

The government-funded facilities available for optical and infrared
astronomy in South Africa are described with an emphasis on the 10-m
Southern African Large Telescope (SALT), due to be commissioned in early
2005. No attempt is made to describe detailed scientific projects or
results, but it is worth noting that productivity remained high throughout
the decade, with the South African Astronomical Observatory (SAAO) annual
report, for example, recording well over 100 publications per year from its
user community. A detailed account of research at the beginning of the
period can be found in the compilation edited by Warner (1995). The
challenge for the future is to redirect these efforts towards making
effective use of SALT.

 Two significant transformations were initiated in South African astronomy
during this decade: one involves the change to big telescope astronomy and
is only just starting; the other is a broad ``Africanization" of activities
which is far from complete but well underway. Prior to 1994 the main
interactions had been with the international community (excluding Africa),
particularly that in the UK. Most of the local optical astronomers had been
born, and many of them trained, outside of Africa. The post-1994 investment
in astronomy came with the assumption that this would change - that
astronomers would find ways to interact with and influence South African
science and that a cohort of indigenous astronomers would be trained and
nurtured. The challenge has been to do this in such a way that these young
scientists are the peers of their international contemporaries and not
merely tokens to fill quotas. Interestingly this challenge is being met
through strengthening collaborations and partnerships, both nationally and
internationally.

\section{National Science and Technology Policy}
 The 1994 government created, among other things, a new Department of Arts,
Culture, Science and Technology (DACST). While there were hopes that these
unlikely bedfellows might encourage a society with a strong culture of
science and technology, the marriage proved unworkable and the 2002 divorce
left the Department of Science and Technology (DST) on its own. DACST (and
latter DST) was a low budget department with a Minister from a minority
party (Inkatha Freedom Party) within the Government of National Unity, which
was led by the African National Congress. Nevertheless, the existence of
DACST, and later DST, as a government department with a specific mandate for
the sciences has had a profound effect on the science policy, and on the
astronomical facilities in particular.
 
In September 1996 DACST issued its first white paper on science and
technology: {\it Preparing for the 21st Century} (DACST 1996). Although
sometimes criticized as not being sufficiently transformational this
document clearly stated the goals then seen as central to South Africa's
young democracy - an improved and sustainable quality of life for all, a
competitive economy and a democratic culture. Considerable emphasis was put
on the challenge to participate in global activities while addressing the
specific needs and aspirations of South Africans. At the core of the new
policy was a vision of a national system of innovation and a clearly
perceived need for capacity development.

In addition to the immediate requirements of society there was an
articulated view of the role of pure science: ``{\it Scientific endeavour is
not purely utilitarian in its objectives and has important associated
cultural and social values. It is also important to maintain a basic
competence in ``flagship'' sciences such as physics and astronomy for
cultural reasons. Not to offer them would be to take a negative view of our
future - the view that we are a second-class nation, chained forever to the
treadmill of feeding and clothing ourselves.}''

In 2002 the newly created Department of Science and Technology published
its {\it Research and Development Strategy} (DST 2002). The main thrusts
have not changed fundamentally: innovation, human resource development for
science and technology and an effective government science and technology
system. Interestingly within the present context, astronomy is specifically
mentioned in several places, e.g. ``{\it One way to achieve national
excellence is to focus our basic science on areas where we are most likely
to succeed because of important natural or knowledge advantages. In South
Africa, such areas include astronomy, human palaeontology and indigenous
knowledge.}" There is also specific mention of the importance of ``{\it
Flagship projects such as the Southern African Large Telescope ...}''

Significantly DST (2002) also points out that South Africa's spending on
science and technology, at 0.7 percent of GDP, is significantly lower than
it should be to ensure national competitiveness in years to come.  Although
there is a commitment from DST to increase the fraction of the GDP that goes
to science, it is not yet clear how this will be achieved.
  
The DST operates through a number of science councils to which it provides
government grants via the National Advisory Council for Innovation. One such
research council, the National Research Foundation (NRF), is directly
involved with astronomy. The NRF provides research funding and evaluation
for tertiary education institutions through its Research and Innovation
Support Agency. It also administers the National Facilities (the largest of
which are physics-based) including SAAO and the Hartebeesthoek Radio
Astronomy Observatory (HartRAO). The NRF and its predecessor, the Foundation
for Research Development (FRD), have been very influential in astronomy in
South Africa during the decade under consideration, due largely to the
direction provided by the FRD/NRF President, Khotso Mokhele, and his vision
for science in Africa\footnote{
www.news.cornell.edu/Chronicle/02/2.28.02/Mokhele.html}.

Resources devoted to higher education are inadequate and the support of
university research by DST and the Department of Education remains a huge
challenge both in terms of retaining talent and in attracting young people
into postgraduate training and academic careers. This is a problem that
should be solved in parallel with the financing of ``flagship" resources.

\begin{figure}
\caption{SALT Ground Breaking Ceremony: Dr Ben Ngubani, Minister of Arts
Culture Science and Technology (left) and Mr Manne Dipico, Northern Cape
Premier, are ``digging". }
\end{figure}

\section{Astronomers in South Africa}
 In reading this account it is worth keeping in mind that the astronomy
community in South Africa is very small - no more than 25 in
optical/infrared astronomy and only about 40 PhDs across the spectrum,
including radio and gamma-ray astronomers and theoretical cosmologists. In
early 2004 most of the active astronomers were based in the two
observatories (National Facilities) with 14 researchers at SAAO and 5 at
HartRAO; the rest were spread thinly among various universities. The number
of research students is disturbingly small, but a strategy is in place to
change this (see section 6.1).

 Ten of South Africa's 21 universities employ astronomers within physics,
mathematics or astronomy departments, but typically only one or two
astronomers per institution. The University of Potchefstroom has a somewhat
larger group engaged in astronomy and space physics, who, among other things,
collaborate in the High Energy Stereoscopic System (H.E.S.S.) gamma-ray
experiment in Namibia\footnote{ www.mpi-hd.mpg.de/hfm/HESS/HESS.html}. The
University of Cape Town has cosmologists, as well as theoretical and optical
astronomers spread amongst their mathematics, physics and astronomy
departments. The University of South Africa, a distance learning
institution, provides the only astronomy undergraduate course in South
Africa. The University of the Free State runs Boyden Observatory with the
recently refurbished 1.5-m (60 inch)
telescope\footnote{www.geocities.com/assabfn/spacetides/boyden.htm} (see
also Paper I).  At the time of writing various tertiary institutions are due
to merge, but it is unlikely that this will significantly affect astronomy.

There are also two planetariums in South Africa. One is part of the Iziko
Museum of Cape Town and the other is run by the University of the
Witwatersrand in Johannesburg. The Astronomical Society of Southern Africa
is an organization of amateur and professional astronomers\footnote{
www.saao.ac.za/assa} who publish an annual astronomical handbook for
Southern Africa (Slotegraaf 2003) and a journal, {\it Monthly Notes of the
Astronomical Society of Southern Africa}.

\section{South African Astronomical Observatory (SAAO)}
 SAAO is the National Facility for optical/infrared astronomy.
Its prime function is to advance fundamental research in astronomy
at a national and international level through the provision and use of a
world-class astronomical facility. The establishment of SAAO was described
in Paper I. SAAO contributes to the development of South Africa by providing
training in a scientific and high-tech environment, by stimulating young
people to follow careers in science and technology through a science
education programme in schools and for teachers and by helping to create a
culture of science and technology amongst all communities through a vigorous
science awareness programme.

Most of the current planning and activity at SAAO revolves directly or
indirectly around SALT. SAAO's {\it vision} for the immediate future 
is {\it to operate SALT as a first-rate international facility with a major
impact on the science and education system of South Africa}. SALT is
described in section 5 and here we look briefly at the other facilities
located at SAAO's Sutherland site, 360 km north-east of Cape Town.

The common-user facility comprises telescopes with apertures of 1.9, 1.0,
0.75 and 0.5m, as well as a 0.75-m automated telescope doing photoelectric
photometry. Although the median seeing, 0.9 arcsec, is not exceptional,
Sutherland is among the darkest and best photometric sites in the world and
SAAO is well known for photometric standards and high precession photometry
of variable stars - at optical and infrared wavelengths. A considerable
variety of techniques are supported, from high speed photometry to long-term
monitoring covering decades. The astrophysics studied encompasses a range of
binary and variable stars including: rapidly oscillating Ap stars, blue
subdwarfs, cataclysmic variables, symbiotic stars, Be stars, asymptotic
giant branch stars, novae, supernovae etc. as well as detailed studies of
Galactic Structure, the Magellanic Clouds and the cosmic distance scale.

An Infrared Survey Facility (IRSF) was opened in November 2000. This 1.4-m
telescope is operated jointly by Japanese (Nagoya and Tokyo Universities)
and South African astronomers.  Its primary mission is a deep survey of the
Magellanic Clouds using {\it SIRIUS}, a simultaneous three-channel imager
operating in the near-infrared $JHK'$ bands, with a $7' \times 7'$
field\footnote{ www.saao.ac.za/facilities/irsf/irsf.html}.

In 2002 YSTAR, a 0.5m automated telescope constructed by Yonsei University,
South Korea, started operations at Sutherland. It performs all-sky
time-series observations in order to identify and monitor various transient
phenomena, including variable stars, near-earth objects and gamma-ray 
bursts\footnote{csaweb.yonsei.ac.kr/$\sim$byun/Ystar/}.

Since 1990 Sutherland has been home to one of the six remote solar
telescopes that make up the Birmingham (UK) Solar Oscillations Network
(BiSON). BiSON is monitoring low-degree oscillation modes of the sun for the
purpose of solar seismology\footnote{bison.ph.bham.ac.uk}.
 
Currently under construction is a 1.2-m telescope; one of two such
instruments that will comprise MONET - a collaboration between the
University of G\"ottingen, SAAO and the McDonald Observatory Texas. The
other telescope is sited next to the Hobby-Eberly telescope in Texas.
Between them the two MONET telescopes cover the entire sky. The programme
has a very large educational component and was funded by the Krupp
Foundation\footnote{ alpha.uni-sw.gwdg.de/$\sim$hessman/MONET/}.

\begin{figure}
\caption{Cut-away diagram showing SALT in its enclosure; the tower houses
the Shack-Hartmann wavefront sensor at the centre of curvature of the
primary mirror array. }
\end{figure}

\section{Southern African Large Telescope (SALT)}
  At the time of writing SALT is nearing completion at Sutherland in the
Northern Cape Province. The telescope will have in essence two primary
missions:\begin{enumerate}
\item {\bf International Mission:} To provide a state-of-the-art facility
that will permit Southern African and international astronomers to perform
cutting-edge research on the southern skies that will complement the
research carried out using telescopes sited in the northern hemisphere.
\item {\bf South-African Mission:} To play a major role in overcoming and
redressing the consequences of the policies of past governments 
that excluded the majority of South Africans from education, training
and careers in science, engineering and technology.
\end{enumerate}

It was very clear by the 1980s that South Africa needed a large telescope
for optical astronomy to be competitive on international terms (Paper I). A
great deal of debate ensued among astronomers as to how large the telescope
should be and of what design, with a minority taking the view that more
small telescopes were to be preferred over a large one. The broad scientific
community, in particular the Royal Society of South Africa and the South
African Institute of Physics, were strongly supportive of astronomy and
appreciated the need for a large telescope (e.g. Ellis 1994).  As apartheid
ended and it became clear that the new government was looking for visionary
projects - a window of opportunity opened.

\subsection{Background to Cabinet Approval}
 In 1996 it was suggested that a copy of the 10-m class Hobby-Eberly
Telescope (HET), then nearing completion at McDonald Observatory in Texas,
would make the ideal large telescope for South Africa. The design of the HET
is particularly cost effective in that it gives access to 70 percent of the
sky for around 20 percent of the cost of a conventional telescope. A 10-m
class telescope had both scientific and political appeal; this model was
adopted by DACST, FRD and the astronomy community who prepared a motivation
to be presented to the South African government (Stobie 1998).

In addition to the strength of astronomy in South Africa at the
time and the very exciting challenges at the
frontiers of scientific knowledge, there were various factors that made a
large telescope attractive to government and/or to international
collaborators.\begin{itemize}
\item {\bf Location:} The latitude and longitude, climate, good
infrastructure, geological stability and the relatively low cost of skilled
labour all contribute to making Sutherland an attractive site for a large
telescope. In addition, Sutherland is located in the Northern Cape, 
a province which is under-resourced in terms of scientific infrastructure.
\item {\bf Focus for competence:} A developing country needs broad
scientific and technical competence, but cannot afford to invest across the
board. Astronomy provides a focus for such an investment in that it draws on
other natural and mathematical sciences, including physics, chemistry,
applied mathematics and computer science. SALT, as a particularly
high-tech telescope, would afford unique technical and engineering training
opportunities.
\item {\bf Science education:} 
International experience has shown that astronomy provides a strong draw
card into science and SALT could provide an inspiration to young and old.
\item {\bf Science Awareness:} A visitor centre associated with SALT would 
attract visitors not only to SALT, but also to Sutherland and to the 
Northern Cape.
\item {\bf African flagship:} There was much to be gained by an endeavour 
in which Africans led, or at least competed on an equal footing with, the 
rest of the world. Astronomy and SALT had the potential to do this at 
modest cost. 
\item {\bf Benefits to industry} Over 50 percent of the cost of the
telescope would be spent within the country providing a stimulation to the
engineering industry. Furthermore, the transfer of cutting-edge technology
from European and North American industry would potentially improve local
competitiveness.
\end{itemize}

Education, particularly science education, is crucial to the success of
South Africa. Black people were systematically excluded from serious science
and mathematics education during apartheid, and the legacy of this is
enormous. Although an obvious area for redress under the new dispensation
progress has been limited - in 2000 less than one percent of the black
Africans taking the school leaving exam passed mathematics at the level
required to proceed with science or engineering studies at university level
(DOE 2001).  There is no doubt that the situation is better than it was in
1994, but given that education is the key to all the national priorities
including health, poverty alleviation and employment, vastly more might have
been done. The issues are too complex to go into in the space available, but
many identifiable problems are identical to, although worse than, those
experienced in North America and Europe, e.g. shortage of properly qualified
teachers. SALT, astronomy and space science will help, but only as part of a
coherent and holistic programme.

On the science awareness front its worth noting the ``impact" of Comet
Shoemaker-Levy 9 (SL9). Contrary to accepted folklore this particular comet
foretold a very positive future for South African astronomy when it crashed
onto Jupiter in 1994. The impact of one fragment of SL9 was televised live
from Sutherland to the world, and drew an extraordinarily large South
African and international audience. One young South African paid it the
ultimate tribute in our sports crazed society: ``{\it The TV coverage ...
was very exciting, a bit like a football match}" (Whitelock \& Flanagan
1998). South Africans, like everyone else, are drawn to science by the
spectacle of astronomy and no better demonstration could have been
orchestrated.
 
When the issue was taken to Cabinet, the SALT project was championed by the
then Deputy Director General for Arts, Culture Science and Technology, Dr Rob
Adam, himself a physicist by training and one of the architects of the
national science policy.

On 1 June 1998 the Minister of Arts, Culture, Science and Technology, the
Hon.\ Lionel Mtshali, announced in Parliament that the South African
Government would fund 50 percent of the total cost of SALT, estimated as
R100M (or US\$20M at the time of the announcement), {\it provided} that the
balance could be raised from international partners. The announcement was
greeted with enthusiastic applause and the decision was strongly supported
by all political parties.

In late 1999, after several international partners expressed strong interest
in contributing to the project, the Cabinet gave SALT the {\it green light}
to proceed, although the telescope was at that stage only about 80 percent
funded. On 1 September 2000 a ground breaking ceremony was held on the SALT
site with representatives from all of the major SALT partners at that time.

The time-table for completion involves commissioning during 2004 and
anticipates operations commencing early in 2005. 

\subsection{International Partnership} 
At the time of writing the SALT consortium comprised 11 partners (including
the HET Board; see section 5.6) from 6 countries, listed here with their
shareholding as of 2002 (the detailed shareholding will change as the
telescope was not fully funded at this stage):\begin{itemize}
\item National Research Foundation (South Africa)\hfill 34.4 \%
\item University of Wisconsin-Madison (USA)\hfill 15.1 \%
\item KBN and CAMK (Poland)\hfill 11.0 \%
\item Rutgers University (USA)\hfill 10.8 \%
\item Dartmouth College (USA)\hfill 9.4 \%
\item G\"ottingen University (Germany)\hfill 4.9 \%
\item United Kingdom SALT Consortium \hfill 4.1 \%
\item University of Canterbury (New Zealand)\hfill 3.9 \% 
\item University of North Carolina (USA) \hfill 3.1 \%
\item Carnegie-Mellon University (funds not yet committed)(USA)\hfill 3.1 \%
\end{itemize}

The total budget for SALT estimated in 1999 US\$ was 30M, divided as 18.12M
for the telescope, 4.8M for instruments and 7.08M for 10 years of
operations. Few of the partners committed to escalating their contributions
with inflation, and as of January 2004 there was a shortfall of between \$2M
and \$3M in respect of the telescope and instruments, for which an additional
partner is being sought.

Furthermore, while the budget for the telescope was probably a reasonable
estimate, that for the operations is certainly an underestimate. The
situation is further complicated by the fact that the budgeting is done in
US\$ while most of the operational costs will be in South African Rand; the
Rand has been very volatile during the last five years.

\subsection{The Telescope}
 The SALT project started as a copy of HET. The only significant change
originally envisaged was a slightly larger altitude tilt, $\rm 37^o$
compared to $\rm 35^o$ for HET, that would give SALT access to the whole of
the Small Magellanic Cloud.  As HET went into operation it became clear that
optimal performance would only be achieved with further modifications and
the SALT project team made a critical systematic analysis of the possible
options. In this regard the project benefited hugely from the experience at
HET and from the insight of those working directly with HET.

 HET and SALT both have primary mirrors comprising 91 identically shaped
spherical mirror segments in a fixed altitude structure. The telescopes
rotate in azimuth to acquire their targets, but remain stationary during an
observation. Compensation for the earth's rotation is achieved by means of a
mobile optical payload which, in the case of SALT, comprises a spherical
aberration compensator and atmospheric dispersion corrector, a moving pupil
baffle and four focal stations including an acquisition camera.

 SALT, its instruments and its progress were described recently by Meiring
et al. (2003) - including the detailed modifications to the HET design - and
by Buckley et al. (2003). The major modifications include:
\begin{itemize}
\item A redesigned spherical aberration corrector (SAC) giving much better 
image quality, a larger field of view ($8'$ diameter {\it vs} $4'$ for HET)
and a larger pupil size (11m {\it vs} 9.2m for HET).
\item Mirror coatings for the SAC that have particularly good ultraviolet 
response ($>90$ percent reflectivity) without compromising the 
visible/infrared performance.
\item Capacitive edge sensors on the primary mirror segments and a 
centre of curvature alignment sensor, that uses a Shack-Hartmann wavefront
sensor, will allow the alignment to be maintained for days at a time and 
will support a possible future phasing of the mirror array.
\item A variety of modifications to minimize dome seeing.
\end{itemize}

\begin{figure}
\caption{SALT Dome; inserts (from bottom left to top right) NASSP students
learn from an expert; unemployed people in Sutherland making eclipse
viewers; children learn optics while constructing their own cardboard
telescopes with plastic lenses.}
\end{figure}
 
\subsection{Instrumentation}
 While the HET design makes it possible to build a low cost telescope, there
is no way to achieve comparable savings on the instrumentation packages
without compromising the overall performance. The intended suite of first
generation instruments included SALTICAM, PFIS and HRS. However, as of early
2004 the design for the High Resolution Spectrograph (HRS) has not been
finalized and its funding is still uncertain. In view of the importance that
the SALT partners place on such an instrument we anticipate good progress by
the time this article is printed.

SALTICAM (O'Donoghue et al. 2003) is the acquisition camera and simple
science imager for SALT. It is mounted at the corrected prime focus and its
wavelength range is 320 to 950 nm. In a simpler form, without re-imaging
optics, it is also being used as the verification instrument to check the
telescope performance during commissioning.

PFIS (Kobulnicky et al. 2003) - the Prime Focus Imaging Spectrograph - is a
versatile high throughput imaging spectrograph operating between 320 and 900
nm. Resolutions of $1500<R<8000$ will be provided by volume-phase
holographic gratings, while the Fabry-Perot mode will allow imaging at
resolutions up to R=12500. Any of the spectroscopic modes can be paired with
the polarimeter to enable linear, circular or all-Stokes polarization
measurements.  It will also be possible to use PFIS at time resolutions
anticipated to be up to 10Hz. An infrared arm to PFIS is likely to be
provided as a second generation instrument.

All of the first generation instruments are being designed to support seeing
limited observations. 

\subsection{Science with SALT}
 Two workshops have been held to discuss possible science projects with SALT
and HET, both in Cape Town (Buckley 1998, 2004). The first of these was held
in 1998 prior to finalization of the telescope and instrument designs and
provided relevant input to the final specifications for the instrumentation.
 
 The second workshop was held in 2003 and attended by more than 100
delegates from 16 countries. Most participants were from SALT partner
institutions, but there was gratifying interest from Africa including 
representatives from Nigeria, Kenya, Uganda, Ethiopia and Mauritius.
Workshops for educators and for students were held in parallel to this
meeting as part of the collateral benefits activities (see section 5.7).

 The possible science projects\footnote{
www.salt.ac.za/content/downloads/stobieworkshops/one/default.htm} discussed
were as far ranging as one might expect, from searching for planets around
nearby stars to surveys of distant quasars. There was consensus that
observers should look for projects in niche areas where SALT might be
particularly efficient, rather than try for direct competition with general
purpose 10-m class telescopes. SALT's special characteristics allow for:
\begin{enumerate}
\item Ultraviolet sensitivity down to the atmospheric cut-off at 320nm;
\item Multi-object spectroscopy over the $8'$ field;
\item Time variability studies: from 0.1s up to one or two hours, or greater 
   than one day;
\item A wide range of polarization studies;
\item High spectral resolution imaging over a few arcmin.
\end{enumerate}
Obviously SALT will not be ideal where high spatial resolution or high 
precision absolute photometry are required.

There had been some hope that the workshop might identify first-light
key-programmes, to which all or most partners would contribute. However, no
consensus could be reached on the desirability for this and no obvious
specific projects were identified. Nevertheless, many interesting ideas were
exchanged and new collaborations were started. There was complete agreement
that rapid proof of SALT's potential to deliver first-rate scientific
results was essential to the telescope's ultimate success. This must be
achieved before serious consideration could be given to funding development
work for the telescope, e.g. for phasing the primary mirror array, or for
the next generation of instruments.

\begin{figure}
\caption{The first of the 91 SALT mirrors being hoisted into position.}
\end{figure}

\subsection{SALT Business Management}
 The SALT Foundation (Pty) Ltd.\ is a South African registered private
company, formed with the purpose of advancing science and training through
the promotion of astronomy in the Southern Hemisphere.  The SALT Foundation
is run through a Board of Directors and the SALT Shareholders Agreement
specifies their relationship with one another and to the company. The
company owns the telescope and partners are shareholders in the company. The
articles of the company specify that its activities shall not be for the
profit of its shareholders and the shareholders enjoy limited liability in
that they are not exposed to the liabilities of the company.

Partners purchase shares in the SALT Foundation (Pty) Ltd. which gives them
access to the telescope in direct proportion to their share. The HET Board,
though not a shareholder, has the right to 10 percent of the observing time
for the first 10 years in exchange for providing HET documentation and
technical assistance.  South Africa, through the National Research
Foundation, is the major participant and shareholder in SALT (see section
5.2) and has two seats on the SALT Board while all other partners have one
each. The first chairperson of the SALT Board was Bob Stobie (Director
SAAO), who died in 2002 and was succeeded by Khotso Mokhele (President of
NRF).

 To the best of our knowledge this is a unique arrangement with respect to
the construction and operation of a telescope. During the construction phase
the SALT Project Manager and the SALT Project Scientist report directly to
the Board. After commissioning is complete the Shareholders' Agreement
specifies that SAAO will operate the telescope under contract to the SALT
Board.

\subsection{SALT Collateral Benefits}
 While SALT is an inexpensive project by international standards, it is a
huge undertaking for South Africa - committing scarce resources that some
might argue were better invested in solving the obvious and debilitating
problems that plague southern Africa, which include health, education,
unemployment and poverty (see also Section 6). The fact that the South
African Government chose to invest in SALT must be seen as an expression of
faith in the long-term prospects for the subcontinent and in the role that
science and technology will play in that future.

 The level of commitment to astronomy is such that much is expected in
return. The SALT Collateral Benefits Plan was established to ensure that
SALT became more than a research tool for a handful of privileged scientists
and that it produces tangible benefits to the people of South Africa. The
plan has five main thrusts: \begin{enumerate}
\item {\bf Industrial empowerment}: aimed at ensuring that South Africa
derives the maximum from the SALT project in terms of technology transfer
and benefits to the economy.
\item {\bf Educational empowerment}:  to provide educational and training          
opportunities, particularly for individuals from historically disadvantaged
communities, during the 5-year construction and 10-plus year operational
phases of SALT.
\item {\bf Public outreach and direct educational empowerment}:  to enhance
science education and awareness throughout South African society,�
inspiring young people to take up careers in science and technology; helping
to create a South African society that is scientifically and technologically
literate.
\item {\bf Science education Visitor Centres}: to develop science and technology        
infrastructure, edu-tourism and educational facilities, particularly in the
Northern Cape.
\item {\bf SALT as an African facility}: to extend the benefits of the space
sciences to the rest of the African continent through collaboration and the
widening of training and scientific development opportunities. These should
allow South Africa and the continent to participate meaningfully in relevant
international scientific endeavours in space and allied sciences.
\end{enumerate}

Developing the collateral benefits in parallel with the telescope and its
instruments has been a major thrust at SAAO over the last five or six years.
Aspects of this are discussed elsewhere in this document and more details
are available on the SALT web page.

It is worth briefly looking at item 5 above and specifically at the name
{\it Southern} African Large Telescope. Clearly South Africa sees its future
as intimately linked with that of the subcontinent. The New Partnership for
African Development (NEPAD) provides a broad strategic framework for
addressing the development needs of the region\footnote{www.nepad.org} and
SALT should be seen within that context - as a potential icon of the African
Renaissance.

Finally, it may be interesting to note that the name - {\it SALT Collateral
Benefits} has proved problematic. This is particularly so in the USA where
the word {\it Collateral} seems to be inseparably linked to {\it Damage}, so
the expression {\it Collateral Benefits} is perceived as an oxymoron, even
worse than {\it Military Intelligence}. Alternatives are under discussion
and the name {\it SALT Science Foundation} has been used, although it is
not ideal.

\section{Strategic Planning and New Developments}
 In February 2001 the National Research Foundation began a process of
strategic planning for astronomy and space science in South Africa with a
meeting attended by all the significant role players from around the
country. This was followed by more intensive discussions within the
Astronomy National Facilities, SAAO and HartRAO, over the course of 2001.
The discussions were far ranging covering possible scientific goals, future
needs in terms of human resources and new equipment, and the broader
requirements of organizational infrastructure.

One factor driving the planning was the fact that SALT, or at least some
sort of large telescope access for South Africa, had been motivated more
than a decade before, but would be operational only in 2005. The time was
ripe to plan for SALT's successor, and perhaps more urgently, to plan a
future for South African radio astronomy.  There was also a clearly
articulated desire on the part of government to see a move away from the
past where isolated centres of excellence interacted with the outside world,
but had relatively little impact on the science system of the country, or
even on each other. In this connection mergers and partnerships were
discussed at length.
 
The following were major outcomes of this process: \begin{enumerate}
\item South African participation in, and bid to host, the Square Kilometre 
Array (SKA)\footnote{www.ska.ac.za}.
\item South African Participation in the World Space Observatory 
(WSO), an ultraviolet space mission\footnote{wso.vilspa.esa.es}.
\item The establishment of the National Astrophysics and Space Science Programme (NASSP).
\item The establishment of an African Institute of Space Science (AISS). 
\item The establishment of an African Virtual Observatory - it is essential
that African astronomers have access to the international virtual
observatories currently being established if they are to use SALT and other
facilities effectively. \end{enumerate} 

The SKA and WSO are outside of the scope of this paper. NASSP and AISS 
are making progress and are discussed briefly below, while South Africa's
links to the Virtual Observatories remains a challenge for the future.

\subsection{National Astrophysics \& Space Science Programme (NASSP)}
 The 2001 strategic planning process identified the lack of appropriately
qualified local scientists, particularly black scientists, as the single
biggest threat to the future of astronomy in South Africa, as well as to the
country at large. The expertise in South African astronomy and space physics
would be quite significant if concentrated in one place rather than spread
thinly through a dozen universities and National Facilities.

NASSP is a training programme that takes students through the honours and
masters degrees\footnote{In SA, students typically do a BSc in 3 yrs, a 1-yr
honours degree and a 2-yr MSc before starting a PhD - the efficacy of this
is debatable and changes are under discussion.}. It is based at the
University of Cape Town (UCT), but the lecturing staff are drawn from the
entire South African astronomy and space physics community, whereby
scientists not located at UCT teach intensively for short periods of time.
The students are drawn from all over the country and from elsewhere in
Africa. Graduates in physics, engineering, mathematics and computer science
are accepted provided they have done physics to the requisite level. One of
the fundamental objectives of NASSP is to create an African network of
astronomers who are bonded by the common experience of schooling and
interlinked both professionally and personally.

NASSP was made possible through the generous sponsorship, initially from the
Ford Foundation and the NRF, and later from the Canon Collins Educational
Trust and the UCT Vice Chancellor's Fund.  Their support made it possible to
pay students realistic bursaries - a crucial factor when poverty is
commonplace among the families of many prospective students.

The first NASSP honours class produced 11 graduates at the end of 2003.
There had been no honours graduates in astronomy anywhere in South Africa
over the previous 3 years and only 2 MSc and 1 PhD graduate over the same 3
year time period. In 2004 15 students have registered through NASSP for
honours and 14 for MSc. This was a very satisfactory start.

NASSP graduates who do not continue in research will go into industry or
commerce - taking with them practical skills in problem solving, data
analysis, computer programming and science communication that will serve
them well in any challenging career.

It is anticipated that many of the NASSP graduates will go on to do PhDs
within South Africa, or with SALT or H.E.S.S. partner institutions who
offer special scholarships for South Africans. Eventually they will obtain
positions in African universities and research establishments.  They will
form the nuclei of research groups who will be users of facilities such as
SALT and H.E.S.S., as well as participants in the broader space science
activities under discussion in Southern Africa.

The NASSP Consortium includes the following universities and National
Facilities:\begin{itemize}
    \item University of Cape Town
    \item University of Natal, Durban and Pietermaritzburg campuses
    \item University of the Free State
    \item Potchefstroom University
    \item University of Zululand
    \item Rhodes University
    \item University of the North West
    \item University of South Africa
    \item South African Astronomical Observatory
    \item Hartebeesthoek Radio Astronomy Observatory
    \item Hermanus Magnetic Observatory
    \end{itemize}
 NASSP is a unique collaboration and it is quite remarkable that such a
diverse group of institutions have been able to compromises and find common
ground with such rapidity. It is a programme very much in the spirit of
NEPAD and the new South Africa and we are delighted that astronomers have
been able to lead in demonstrating how collaborations can benefit all
involved - the whole is vastly greater than the sum of the parts.

More details of NASSP can be found on its website\footnote
{www.star.ac.za}.

\begin{figure}
\caption{Sutherland landscape; inserts: (left to right) a
young technician trains in the mechanical workshop; SALT mirror support
structure with central 7 mirrors and NASSP students; student fills cryostat
with liquid nitrogen.}
\end{figure}

\subsection{African Institute of Space Science (AISS)}
 The concept of an African Institute of Space Science (AISS) was developed
as a way to mobilize the space sciences, including astronomy, in support of
the national development agenda. Through their multi-disciplinary and
cross-disciplinary nature, the space sciences are in a strong position to
capitalize innovation and technological development. AISS will combine
various activities around southern Africa to provide a coordinated strategic
programme for developing space science in the region. Among other things,
AISS can provide a local point of contact for international linkages to
southern African space science. By forging partnerships of South African
government departments, academia and industry with similar agencies
internationally, AISS will be in a strong position to leverage funds and to
spearhead appropriate applications of space science and technology to
African development issues.

The AISS initiative has been embraced by the NRF and has attracted
considerable interest from African space scientists (e.g. UN 2001). It has
brought to light the current lack of coordination and consequently the
rather disparate nature of South Africa's space science and technology
activities that are supported by government. The NRF continues to pursue the
AISS initiative through a newly established inter-departmental National
Working Group on Space Science and
Technology\footnote{www.nrf.ac.za/publications/news@nrf/feb2002/africaspace.stm}
which comprises representatives from the Departments of Communication, of
Trade and Industry, and for Administration in addition to DST and a number
of other partners in the SA space arena.

\section{Conclusion}
 I want to finish by contextualizing the spending on astronomy in South
Africa through comparing expenditure on SALT with that on the Hubble Space
Telescope (HST). The cost of HST at launch in 1990 was \$1.5 billion
(or \$1.59 billion in 1992 assuming 3 percent inflation) while that of
SALT in 2002 terms is about \$25 million. Using data from the
World Bank\footnote{www.worldbank.org/data/countrydata.html} we see that the
USA GDP in 1992 was \$6.262 trillion, while that of South Africa in 2002 was
\$104.2 billion. The cost of SALT compared to the GDP of South Africa, 0.025
percent, is almost identical to that of HST compared to the GDP of the USA.
There can be no doubt that the South African astronomy community is
extraordinarily fortunate in having access to this level of support.

 The politicization of science is often problematic for scientists, and much
has been written about the inevitability of this process in connection with 
big science projects (e.g. Enard 2002). The above comparison with HST
demonstrates that SALT is big science for Africa; the people of South
Africa will have expectations of it that are comparable to those of
Americans for HST. We South Africans will do well to follow the example of
the Space Telescope Science Institute in ensuring that: \begin{itemize}
\item observing time goes to the astronomers with the best projects,
\item those astronomers are fully empowered to do first rate science with 
SALT 
\item and that the outcomes of their research are made accessible to the
public and particularly to young Africans.\end{itemize}

 It seems appropriate to give the last word to the President of
South Africa, Thabo Mbeki. In opening the South African Pavilion at the
2000 World Expo in Hanover, Germany, President Mbeki said the following:
\begin{quotation}``{\it
Now, in the small town of Sutherland in the semi-desert Karoo region of our
country, we are building a gigantic African eye through which we can view
the universe. The construction of the single largest telescope in the
southern hemisphere, SALT - as it is called - will mean that in this humble
home of our earliest humans, we are also building a vast gateway through
which we can observe our earliest stars, learn about the formation of our
galaxy and the lives of other worlds so as to give us insights into our
future.  We are proud that SALT will not only enable South African
scientists to undertake important research, but also provide significant
opportunities for international collaboration and scientific partnerships
with the rest of the world.}"\end{quotation}

\acknowledgements My thanks to Shireen Davies for her help with the
illustrations and the following for their advice and critical reading of
this manuscript - Luis Balona, David Buckley, Michael Feast, Peter Martinez,
Kobus Meiring, John Menzies, Clifford Nxomani and Darragh O'Donoghue.

\end{document}